\documentstyle[preprint,revtex]{aps}
\tightenlines
\begin{document}
\draft
\begin{title}
Off-shell Behavior of the $\pi\!-\!\eta$ Mixing Amplitude\\
\end{title}
\author{J. Piekarewicz}
\begin{instit}
Supercomputer Computations Research Institute, \\
Florida State University, Tallahassee, FL 32306
\end{instit}

\begin{abstract}
We extend a recent calculation of the momentum dependence
of the $\rho-\omega$ mixing amplitude to the pseudoscalar
sector. The $\pi\!-\!\eta$ mixing amplitude is calculated in a
hadronic model where the mixing is driven by the neutron-proton
mass difference. Closed-form analytic expressions are presented in
terms of a few nucleon-meson parameters. The observed momentum
dependence of the mixing amplitude is strong enough as to question
earlier calculations of charge-symmetry-breaking observables based on
the on-shell assumption. The momentum dependence of the $\pi\!-\!\eta$
amplitude is, however, practically identical to the one recently
predicted for $\rho-\omega$ mixing. Hence, in this model, the
ratio of pseudoscalar to vector mixing amplitudes is, to a good
approximation, a constant solely determined from nucleon-meson coupling
constants. Furthermore, by selecting these parameters in accordance with
charge-symmetry-conserving data and SU(3)-flavor symmetry, we reproduce
the momentum dependence of the $\pi\!-\!\eta$ mixing amplitude predicted
from chiral perturbation theory. Alternatively, one can use
chiral-perturbation-theory results to set stringent limits on the value
of the $NN\eta$ coupling constant.
\end{abstract}

\narrowtext

\section{Introduction}
\label{secintro}

Most theoretical efforts devoted to the understanding of meson mixing in
charge-symmetry-violating (CSV) observables start from a nucleon-nucleon
($NN$) interaction constructed from coupling constants previously determined
from empirical two-nucleon data. In addition, meson mixing amplitudes are
obtained, either, from experiment (in the case of $\rho-\omega$ mixing) or
are inferred (for $\pi\!-\!\eta$ mixing) from the mass splittings of the
SU(3) octet of pseudoscalar mesons~\cite{coon75,coon82}. These values for
the mixing amplitudes, however, reflect physics relevant to the
timelike region. Nevertheless, these on-shell values are still used in
constructing the CSV component of the $NN$ interaction which requires,
instead, information about meson mixing in the spacelike region.

In a recent publication, Goldman, Henderson, and Thomas have questioned
the previously accepted procedure of using the on-shell value for the
$\rho-\omega$ mixing amplitude in calculating CSV
observables~\cite{goldman91}. Two calculations that address the momentum
dependence of the $\rho-\omega$ mixing amplitude have recently been
completed. In the first calculation, Goldman, Henderson, and Thomas, have
estimated the momentum dependence of the mixing amplitude in a model where
the mixing was generated by quark-antiquark ($q\bar{q}$) loops and thus
driven by the up-down quark mass difference~\cite{goldman91}.

More recently, Piekarewicz and Williams have also tested the on-shell
prescription in a model in which the mixing was, in contrast, generated
by nucleon-antinucleon ($N\bar{N}$) loops and hence driven by the
neutron-proton mass difference~\cite{piewil93}. Closed-form analytic
expressions for the mixing amplitude were presented in terms of a few
nucleon-meson parameters. An important assumption of their model is that
the coupling of mesons to $N\bar{N}$ loops is determined by the underlying
theory and, therefore, ultimately constrained by empirical two-nucleon
(NN) data. Using standard values for these couplings they reported a value
for the $\rho-\omega$ mixing amplitude at the on-shell point in agreement
with experiment.

In spite of some obvious differences between the two models, both
calculations predicted a substantial momentum dependence for the
$\rho-\omega$ mixing amplitude and suggested that the presence of a
node in the $NN$ potential around $r\sim 0.9$~fm should suppress the
$\rho-\omega$ contribution to the CSV potential~\cite{goldman91,piewil93}.
To lend support to these claims, Maltman and Goldman have tested
(in a calculation of $\pi\!-\!\eta$ mixing) the assumptions underlying
the quark-loop model by using constraints imposed from chiral perturbation
theory~\cite{maltman92,maltgol92}. They concluded that the quark-loop model
of Ref.~\cite{goldman91} should provide a reliable estimate of the momentum
dependence of the (off-diagonal) $\rho-\omega$ mixing amplitude at least in
the region of applicability of chiral perturbation theory.

These findings suggest that previous calculations of CSV observables
based on the on-shell assumption are suspect and should be re-examined.
For example, $\rho-\omega$ mixing is believed to account for half of
the size of the difference between the neutron and proton analyzing
power measured in elastic neutron-proton scattering at
IUCF~\cite{knutsen90,miller86,willia87}. Furthermore, $\rho-\omega$
mixing seems to also play an important role in explaining
the binding-energy difference between mirror nuclei (Nolen-Schiffer
anomaly)~\cite{nolsch69} as well as in accounting for the difference
in $NN$ scattering lengths~\cite{coon75,coon87,bluiqb87}.

However, it might be difficult to reexamine these results by looking at
$\rho-\omega$ mixing in isolation. Although $\rho-\omega$ mixing has been
recognized as an important source of charge symmetry violation, its role
in describing observables is far from preeminent. Indeed, most CSV
observables arise from a, sometimes delicate, sum of many different
contributions~\cite{miller90}. Furthermore, some of these additional
contributions, specifically $\pi\!-\!\eta$ mixing, have been calculated
by employing the now suspect on-shell approximation. It is therefore
natural, in view of the substantial momentum dependence displayed by
the $\rho-\omega$ mixing amplitude, to extend our calculations to
the pseudoscalar sector.

Pseudoscalar mixing has traditionally been regarded as marginally important
in understanding CSV observables. For example, $\pi\!-\!\eta$ mixing does not
contribute to the neutron-proton analyzing power difference. Furthermore,
$\pi\!-\!\eta$ mixing is believed to generate a small contribution
(of the order of 10\%) to the difference in $NN$ scattering lengths
as well as to the Nolen-Schiffer anomaly~\cite{coon75,coon87,bluiqb87}.
This should be contrasted with vector ($\rho-\omega$) mixing which, until
very recently, was believed to play a crucial role in explaining
all three CSV observables. One should realize, however, that once the
on-shell prescription is relaxed and the full momentum dependence of the
mixing amplitudes is incorporated, it is not clear what the final outcome
of the calculations will be.

In this work we extend our previous hadronic calculation of vector
mixing to the pseudoscalar sector. We assume that the $\pi\!-\!\eta$
mixing amplitude is generated by $N\bar{N}$ loops and thus driven by the
neutron-proton mass difference. In Sec.~{\ref{sectwo}} we present the
formalism and evaluate the mixing amplitude analytically. As in the case
of vector ($\rho-\omega$) mixing we obtain a strongly momentum-dependent
$\pi\!-\!\eta$ mixing amplitude. In spite of the substantial momentum
dependence displayed by both mixing amplitudes we find an almost constant
ratio of pseudoscalar to vector amplitudes. In Sec.~{\ref{secthree}}
we show that by employing accepted values for the nucleon-meson parameters
we are able to reproduce the momentum dependence of the $\pi\!-\!\eta$
mixing amplitude predicted from a recent calculation using chiral
perturbation theory ({\raise 2pt \hbox{$\chi$}}PT)~\cite{maltman92}.
Finally, our conclusions and suggestions for future work are presented
in Sec.~{\ref{secfour}}.

\section{Formalism}
\label{sectwo}

The starting point for our calculation is a hadronic model
in which the $NN$ interaction is mediated by several meson exchanges.
For the purpose of this calculation it is sufficient to specify the
coupling of nucleons to the two lightest pseudoscalar mesons. For
the coupling of nucleons to the neutral pion we will assume a
pseudoscalar representation
  \begin{equation}
    {\cal L}_{{\scriptscriptstyle NN}\pi}=-i
     g_{\pi}\bar{\psi}\gamma^{5}
     \vec{\tau}\cdot\psi\vec{\pi} \;.
   \label{lpi}
  \end{equation}
Although there are reasons to believe that a pseudovector coupling
might be more appropriate in calculating low-energy
observables~\cite{matser82,serwal86}, we will show, at least for a
mixing amplitude calculated to one loop, that the two representations
yield an identical momentum dependence. Because of the special role played
by the pion in the $NN$ interaction, the value of the $NN\pi$ coupling
constant is very well known. In contrast, the $NN\eta$ coupling is not
well determined by one boson exchange (OBE) models of the $NN$ interaction.
In particular, the full Bonn potential yields excellent fits to
two-nucleon data without introducing a pseudoscalar-isoscalar
meson~\cite{machl86,machl87}. The $NN\eta$ coupling constant is,
instead, determined from SU(3)-flavor symmetry and is conventionally
assumed to vary in the range $g_{\eta}^{2}/4\pi\sim
(0.50-1.00)$~~\cite{coon75,coon82,henmil79}. In the present work we use
a pseudoscalar representation for the
$NN\eta$ coupling
  \begin{equation}
    {\cal L}_{{\scriptscriptstyle NN}\eta}=-i
    g_{\eta}\bar{\psi}\gamma^{5}\psi{\eta} \;,
   \label{leta}
  \end{equation}
and adopt the value of Ref.~\cite{coon82} for the $NN\eta$ coupling
constant (see Table~\ref{tableone}).

Given the interaction Lagrangian, one can then proceed to
calculate the contribution from $\pi\!-\!\eta$ mixing to the
$NN$ potential~\cite{coon75,henmil79},
  \begin{equation}
    \hat{V}^{\pi\eta}_{III}(q) = -
     {g_{\pi}g_{\eta}
     \langle \pi | H | \,\eta \rangle \over
     (q^{2}-m_{\pi}^{2})
     (q^{2}-m_{\eta}^{2})} \,
       \gamma^{5}(1)\,\gamma^{5}(2)
       \Big[\tau_{\scriptscriptstyle z}(1)+
            \tau_{\scriptscriptstyle z}(2)\Big] \;.
     \label{vthree}
  \end{equation}
In the present work we employ the same hadronic model previously used
in calculating $\rho-\omega$ mixing to evaluate the momentum dependence
of the $\pi\!-\!\eta$ mixing amplitude. Most of the formalism has been
presented before and we include some of the details here just for
completeness~\cite{piewil93}.

The $\pi\!-\!\eta$ mixing amplitude can be evaluated to leading
(one-loop) order
  \begin{equation}
     \langle \pi | H | \,\eta \rangle_{\rm ps} = -
     g_{\pi}\,g_{\eta}\,\Pi_{{\rm ps}}(q^2) \;,
   \label{hpietaps}
  \end{equation}
in terms of the pion-eta mixing self-energy
  \begin{equation}
     i\Pi_{\rm {ps}}(q^2) =
     \int {d^4k\over (2\pi)^4} {\rm Tr}
     \left[
       \gamma^{5} \tau_{\scriptscriptstyle z} G(k+q)
       \gamma^{5} G(k)
      \right] \;,
      \label{pips}
   \end{equation}
with an assumed pseudoscalar $NN\pi$ coupling. The isospin trace can be
evaluated by writing isoscalar and isovector components of the nucleon
propagator
  \begin{equation}
G(k)={1\over 2}G_{p}(k)(1+\tau_{\scriptscriptstyle z}) +
     {1\over 2}G_{n}(k)(1-\tau_{\scriptscriptstyle z}) \equiv
     G_{0}(k)+G_{1}(k)\tau_{\scriptscriptstyle z}
  \end{equation}
in terms of individual proton and neutron contributions
  \begin{equation}
	G_{p}(k)=
         {\rlap/{k} + M_{p} \over
          k^2-M_{p}^{2}+i\epsilon} \;, \quad
	G_{n}(k)=
         {\rlap/{k} + M_{n} \over
          k^2-M_{n}^{2}+i\epsilon} \;.
   \label{greens}
  \end{equation}
After performing the isospin trace one obtains a $\pi\!-\!\eta$ mixing
amplitude driven by the difference between proton and neutron loops:
  \begin{equation}
     \Pi_{{\rm ps}}(q^2) =
     \Pi_{{\rm ps}}^{(p)}(q^2) - \Pi_{{\rm ps}}^{(n)}(q^2) \;.
   \label{pipsdiff}
  \end{equation}
As is common place in field-theoretical models, most of the
integrals appearing in calculating vacuum correction to tree-level
amplitudes are divergent and must be renormalized. To isolate the
singularities one first regularizes all integrals and then removes
the divergences by inserting appropriate counterterm contributions.
For example, one can isolate the singularities in the proton contribution
to the $\pi\!-\!\eta$ mixing amplitude by using dimensional
regularization~\cite{ramond81},
   \begin{eqnarray}
     \Pi_{{\rm ps}}^{(p)}(q^2) = &-&
     {1 \over 4\pi^2}
     \Bigg[
      \Gamma(\epsilon)
      \left( M_{p}^{2}-{q^{2} \over 2} \right) +
      {q^{2} \over 3} \nonumber \\ &-&
      \int_{0}^{1} dx
      \left( M_{p}^{2}-3x(1-x)q^{2} \right)
      \ln \left(M_{p}^{2}-x(1-x)q^{2} \over \Lambda^{2}\right)
     \Bigg] \;.
     \label{pipsinf}
   \end{eqnarray}
Here $\Lambda$ is an arbitrary renormalization scale and
$\Gamma(\epsilon)=(\epsilon^{-1}-\gamma+\cdots)$ is the gamma
function evaluated in the limit of $\epsilon \rightarrow 0$,
and $\gamma$ is the Euler-Mascheroni constant. Having
isolated the singularities one can render the integral finite
by appropriate counterterm subtractions. Notice, however, that
in contrast to the $\rho-\omega$ mixing amplitude, the singularities
can not be removed by simply subtracting the corresponding neutron
contribution~\cite{piewil93} [note that the singularity is now
proportional to the mass term; see Eq.~(\ref{pipsinf})].
Nevertheless, all divergences can be
eliminated by appropriate counterterm subtractions~\cite{matser82,serwal86}.
The precise value for the (infinite) coefficients must be specified
by imposing appropriate renormalization conditions. Since the
proton-neutron subtraction removes the $q^2$ singularity a single
counterterm is sufficient to render the amplitude finite. This
counterterm is chosen in such a way that the $\pi\!-\!\eta$ mixing
amplitude reproduces the ``experimental'' value at the on-shell
$\eta$-meson point~\cite{coon75,coon82}:
  \begin{equation}
     {\langle \pi | H | \, \eta \rangle}_{\rm ps}
     \Big|_{q^2=m_{\scriptscriptstyle \eta}^{\scriptscriptstyle 2}}=
     -4200\,{\rm MeV}^{2} \;.
   \label{pionshell}
  \end{equation}
Hence, after performing the appropriate subtractions one obtains a
finite $\pi\!-\!\eta$ mixing amplitude that can be written as
   \begin{equation}
     \Pi_{{\rm ps}}(q^2)   \equiv
     \Pi_{{\rm ps}}(q^2\!=\!0) + \Pi_{{\rm pv}}(q^2) \;,
    \label{pipsfin}
   \end{equation}
where
   \begin{equation}
     \Pi_{{\rm pv}}(q^2) = -
      {q^2 \over 8\pi^2}
      \int_{0}^{1} dx
      \ln \left[
        {M_{p}^{2}-x(1-x)q^{2} \over M_{n}^{2}-x(1-x)q^{2}}
     \right] \;,
    \label{pipvfin}
   \end{equation}
and where the constant $\Pi_{{\rm ps}}(q^2\!=\!0)$ is adjusted so that
the renormalization condition [Eq.~(\ref{pionshell})] is fulfilled.
The full momentum dependence of the $\pi\!-\!\eta$ mixing amplitude is,
therefore, contained in the term $\Pi_{{\rm pv}}(q^2)$. This expression
can be evaluated in closed form and reduces, to leading order in the
neutron-proton mass difference, to the following simple form:
  \begin{eqnarray}
     {\Pi_{\rm pv}(q^2) \over M^2} =
     {1 \over \pi^2}
     {\Delta M \over M}
     \cases{
       \displaystyle{1\over \xi}
       \,{\rm tan}^{-1}
       \left( \displaystyle {1 \over \xi} \right) \,, &
       for $0 < q^2 < 4M^2 \;;$ \cr
       ${\phantom {xxx}}$  &   \cr
       \displaystyle{1\over 2\xi}
       \ln \left( \left|
        \displaystyle{\xi-1 \over \xi+1}
       \right| \right) \,, &
       ${\rm otherwise} \;,$ \cr}
   \label{pietaan}
  \end{eqnarray}
where
  \begin{equation}
   M = {1\over 2}(M_{n}+M_{p}) \;, \quad
   \Delta M= (M_{n}-M_{p})   \;, \ {\rm and} \quad
   \xi=\displaystyle{
    \left| 1-{4M^2 \over q^2} \right|^{1/2}} \;.
  \end{equation}

We have appended the pseudovector (pv) subscript to the above expression
because it is, in fact, the result that one would have obtained if a
pseudovector (as opposed to a pseudoscalar) $NN\pi$ coupling had been
adopted. Hence, pseudoscalar and pseudovector representations generate
the same momentum dependence for the $\pi\!-\!\eta$ mixing amplitude.
The equivalence (at least to one loop) between pseudoscalar and
pseudovector representations is not difficult to understand.
For on-shell nucleons the equivalence of the two representations is well
known. Since the imaginary part of vacuum polarization is related to the
decay of a pseudoscalar meson into (on-shell) $N\bar{N}$ pairs, both
representations yield identical imaginary parts~\cite{serwal86}.
Moreover, the polarization is an analytic function of $q^{2}$. Hence,
the real part of the polarization can be written in terms of a (subtracted)
dispersion integral involving only the imaginary part. Consequently, both
representations should (and do) yield, up to a constant, the same one-loop
mixing amplitude.

	As we will show later, vacuum corrections generate a substantial
momentum dependence for the $\pi\!-\!\eta$ mixing amplitude (see
Fig.~\ref{figone}). By itself, this should be sufficient reason to
question earlier results based on the on-shell assumption. However,
one should first test the reliability of the model. For example,
could it be possible to calculate other observables ({\it e.g.,} ratio
of mixing amplitudes) that might be less sensitive to the model assumptions?
Can one compare these hadronic results with other calculations ({\it e.g.,}
chiral perturbation theory) that, while having a limited ($q^2$) range
of applicability, might be perceived as having a stronger theoretical
underpinning? (See Ref.~\cite{maltgol92} for a comparison of quark-loop
model results to predictions from chiral perturbation theory). In what
follows we will try to answer some of these questions.

In a recent work on $\rho-\omega$ mixing we showed that the mixing
amplitude calculated in a hadronic model was given by
  \begin{equation}
     \langle \rho | H | \,\omega \rangle =
     g_{\rho}\,g_{\omega}\,q^2\,
     \left[\Pi_{\rm vv}(q^2)+C_{\rho}\Pi_{\rm vt}(q^2)\right] \;,
   \label{hrhoom}
  \end{equation}
where $g_{\rho}(g_{\omega})$ is the $NN\rho$($\omega$) coupling
constant, $C_{\rho}\equiv f_{\rho}/g_{\rho}$ is the ratio of tensor
to vector $NN\rho$ coupling (see Table~\ref{tableone}), and
$\Pi_{\rm vv}$ and $\Pi_{\rm vt}$ are the transverse components of
the vector-vector and vector-tensor polarizations
respectively~\cite{piewil93}.

      The existence of a nonzero $\rho-\omega$ mixing amplitude is,
of course, very well established. The mixing has been experimentally
observed in the measurement of the pion form factor at the $\omega$-meson
point~\cite{miller90,henmil79,barkov85}. In contrast, no direct
experimental measurement exists for $\pi\!-\!\eta$ mixing. The magnitude
of the mixing must, therefore, be inferred from the mass splitting of
the pseudoscalar octet~\cite{coon75,coon82}. However, the theoretical
procedure employed in extracting the value of the $\pi\!-\!\eta$ mixing
amplitude is now the subject of some controversy~\cite{maltman92}. Thus,
it is safe to assume that the on-shell value of the $\pi\!-\!\eta$ mixing
amplitude is not well known much less its full momentum dependence.

	The momentum dependence of the $\pi\!-\!\eta$ mixing amplitude,
however, can be constrained, at least within the present hadronic model,
from knowledge of the of the off-shell behavior of the corresponding
vector amplitude. Indeed, the pseudoscalar and vector mixing amplitudes
are intimately related in the model. The origin for this relation
is the following identity between vacuum polarization amplitudes:
  \begin{equation}
   \Pi_{\rm pv}(q^2)=-q^2\Pi_{\rm vt}(q^2) \;.
   \label{piident}
  \end{equation}
This, in turn, translates into the following relation between the
ratio of mixing amplitudes
  \begin{equation}
    {\langle \pi  | H | \,\eta   \rangle_{\rm pv} \over
     \langle \rho | H | \,\omega \rangle_{\phantom{\rm pv}}} =
     \left(
       {g_{\pi} \,g_{\eta}  \over
        f_{\rho}\,g_{\omega}}
     \right)
     \left[
        1 \over
        1 + g_{\rho}\Pi_{\rm vv}/f_{\rho}\Pi_{\rm vt}
     \right] \;,
    \label{piratio}
   \end{equation}
where $\langle \pi  | H | \,\eta  \rangle_{\rm pv}$ is the
$\pi\!-\!\eta$ mixing amplitude minus its value at $q^2=0$.
Because of the similar momentum dependence of $\Pi_{\rm vv}$ and
$\Pi_{\rm vt}$, the above relation indicates that the ratio of
mixing amplitudes is essentially constant. We believe that such
relations between mixing amplitudes might be useful in the study
of meson mixing, particularly, if they prove to be less sensitive
to the assumptions of the model (notice that the weak momentum dependence
of the ratio will be preserved even after the inclusion of form factors;
see Fig.~\ref{figtwo}).

\section{Results}
\label{secthree}

	The momentum dependence of the $\pi\!-\!\eta$ mixing amplitude is
shown in Fig.~{\ref{figone}}. Two sets of calculations are displayed.
One set (solid and dashed lines) uses a value of $g_{\eta}^{2}/4\pi=0.5$
for the $NN\eta$ coupling constant~\cite{coon75}. The other set (dot-dashed
and dotted lines) uses, instead, the more recent value of
$g_{\eta}^{2}/4\pi=0.9$~\cite{coon82}. Furthermore, the solid and
dot-dashed lines show results for the mixing amplitude using point
nucleon-meson couplings at all values of $q^{2}$. In contrast, the dashed
and dotted lines show results for a mixing amplitude modified by the
introduction of form factors in the spacelike region. These form factors
are introduced by modifying the point coupling in the following
way:
  \begin{equation}
    g_{\pi} \rightarrow  g_{\pi}(q^2) \equiv
    g_{\pi} \left(
    1-q^{2}/\Lambda_{\pi}^{2} \right)^{-1} \;; \quad
    g_{\eta} \rightarrow  g_{\eta}(q^2) \equiv
    g_{\eta} \left(
    1-q^{2}/\Lambda_{\eta}^{2} \right)^{-1} \;, \quad
   \label{formf}
  \end{equation}
with the numerical values for the cutoffs determined from empirical
two-nucleon data (see Table~{\ref{tableone}}). Also shown (using
data-like symbols) are recent results obtained by Maltman using
chiral perturbation theory~\cite{maltman92}. The size of the
``error bars'' is meant to represent the spread in the results
induced by the uncertainty in the theoretical determination of the
electromagnetic contribution to the mass-squared splitting in the
kaon system. Although the overall normalization of the hadronic
result was obtained by imposing appropriate renormalization conditions,
the fact that the observed momentum dependence agrees with the predictions
from chiral perturbation theory is a significant result.
This finding, added to the success of the hadronic model in reproducing
the value of the $\rho-\omega$ mixing amplitude at the on-shell
point~\cite{piewil93}, gives us confidence in extending the model
beyond the region of applicability of chiral perturbation theory.

Perhaps the most uncertain parameter in our calculations is the value
of the $NN\eta$ coupling constant (the remaining three parameters,
$g_{\pi}, M_{p},$ and $M_{n}$, are very well known). Hence, one could
attempt to use results from {\raise 2pt \hbox{$\chi$}}PT to set
some limits on the hadronic $NN\eta$ coupling constant.
In fact, the constraints imposed by {\raise 2pt \hbox{$\chi$}}PT
on $g_{\eta}$ are tighter than Fig.~\ref{figone} might suggest.
This can be observed by expanding our analytic results for the
$\pi\!-\!\eta$ mixing amplitude [Eq.~(\ref{pietaan})] to leading order
in $q^2/4M^{2}$, {\it i.e.,}
  \begin{equation}
     {\langle \pi | H | \, \eta \rangle}_{\rm ps}  \simeq -
     \left[ a_{0} + a_{1}q^{2}/m_{\eta}^{2} \right] \;,
  \end{equation}
where
  \begin{equation}
    a_{0} =  g_{\pi}g_{\eta} \Pi_{{\rm ps}}(q^2\!=\!0) \;; \qquad
    a_{1} = {g_{\pi}g_{\eta} \over 4\pi^{2}}
                 \left( {\Delta M \over M} \right)
                 m_{\eta}^{2} \;.
  \end{equation}
The value for the constant $\Pi_{{\rm ps}}(q^2\!=\!0)$ (and hence
$a_{0}$) has been chosen to satisfy the on-shell renormalization
condition [see Eq.~(\ref{pionshell})]. The value for the slope,
on the other hand, in conjunction with the following limits set by chiral
perturbation theory~\cite{maltman92}
   \begin{eqnarray*}
         280 \,{\rm MeV}^{2} \le a_{1} \le 360 \,{\rm MeV}^{2} \;,
   \end{eqnarray*}
can be used to constrain the value of the $NN\eta$ coupling constant to
the range
   \begin{eqnarray*}
         0.32 \le {g_{\eta}^{2} \over 4\pi} \le 0.53 \;.
   \end{eqnarray*}
These values are somehow smaller, but still consistent, with the
limits \hbox{$g_{\eta}^{2}/4\pi\sim$} \hbox{$(0.5-1.0)$} inferred
from SU(3)-flavor
symmetry~\cite{coon75,coon82,henmil79}. Furthermore, these results
support the assertion of Ref.~\cite{coon82} that a value of
$g_{\eta}^{2}/4\pi \sim 0.9$ does, indeed, provide an
upper limit on the importance of pseudoscalar mixing and that values
as large as $g_{\eta}^{2}/4\pi \sim 4$, as seems to be required by
static approximations to OBE models of the NN
interaction~\cite{machl86,machl87}, are not realistic.

	The ratio of the pseudoscalar to the vector mixing amplitude
[Eq.~(\ref{piratio})] is plotted in Fig.~\ref{figtwo} with (dashed
line) and without (solid line) form factor corrections. This
result indicates that in spite of the substantial momentum dependence
displayed by both amplitudes, their ratio is essentially constant over
the entire range of momentum shown in the figure. Notice that this result
holds even in the presence of form factor corrections (variations are less
than 10\%). Furthermore, because the tensor to vector $NN\rho$ coupling
is large ($C_{\rho}=6.1$), and because $\Pi_{\rm vv} \sim \Pi_{\rm vt}$,
the ratio of pseudoscalar to vector mixing amplitudes can be approximated
by a simple ratio of nucleon-meson coupling constants
  \begin{equation}
    {\langle \pi  | H | \,\eta   \rangle_{\rm pv} \over
     \langle \rho | H | \,\omega \rangle_{\phantom{\rm pv}}} \simeq
     \left(
       {g_{\pi} \,g_{\eta}  \over
        f_{\rho}\,g_{\omega}}
     \right) \;.
    \label{piratiob}
   \end{equation}

Finally, in Fig.~\ref{figthree} we present the contribution from
$\pi\!-\!\eta$ mixing to the singlet (${}^{1}\!S_{0}$) component of the NN
interaction in the nonrelativistic limit. This is given, for example, in the
on-shell limit and with no form factor corrections by~\cite{coon75}
  \begin{equation}
     V^{\pi\eta}_{{}^{1}\!S_{0}}(q) = -
     {g_{\pi}g_{\eta}
     \langle \pi | H | \,\eta \rangle_{\rm ps} \over
     ({\bf q}^{2}+m_{\pi}^{2})
     ({\bf q}^{2}+m_{\eta}^{2})} \,
     \left({{\bf q}^{2} \over 4M^{2}}\right) \;.
     \label{vsinglet}
  \end{equation}
This result is indicated by the solid line. The dashed line shows
results obtained using also on-shell mixing but with the point
coupling modified by form factor corrections according to
Eq.~(\ref{formf}). Finally, the dot-dashed line shows the effect of
off-shell mixing and form factors on the potential. Although off-shell
mixing reduces the overall strength of the potential, the size of the
suppression is comparable to the reduction observed in going from point
couplings to form factors. Consequently, these changes are not as dramatic
as the ones observed in the analogous calculation of $\rho-\omega$
mixing~\cite{piewil93}. There, in addition to a strong suppression,
off-shell mixing generated a NN potential opposite in sign (over the
entire spacelike region) relative to the conventional on-shell contribution.

\section{Conclusions}
\label{secfour}

	We have calculated $\pi\!-\!\eta$ mixing in a hadronic model
where the mixing was generated by $N\bar{N}$ loops and thus driven
by the neutron-proton mass difference. We have presented closed-form
analytic expressions for the mixing amplitude in terms of a few
meson-nucleon parameters. By adopting values for these parameters
inferred from SU(3)-flavor symmetry or from CSC two-nucleon data,
we obtained a momentum dependence for the mixing amplitude in
agreement with recent results from chiral perturbation theory.
Alternatively, {\raise 2pt \hbox{$\chi$}}PT results were used to
constrain the value of the hadronic $NN\eta$ coupling constant to
the range $0.32 \le {g_{\eta}^{2} / 4\pi} \le 0.53$. These values
were slightly smaller, albeit still consistent, with the limits inferred
from SU(3)-flavor symmetry.

	We have, also, related the off-shell behavior of the
$\pi\!-\!\eta$ and $\rho-\omega$ mixing amplitudes. In the present model
the momentum dependence of the pseudoscalar and vector amplitudes are
not independent. In fact, we found the ratio of pseudoscalar to vector
mixing amplitudes to be practically constant and determined essentially
from meson-nucleon coupling constants. Since this result was only slightly
modified by form factor corrections, we believe that the ratio might be
less sensitive to the particular assumptions of the model.

	To gauge the effects from off-shell mixing, we calculated the
contribution from $\pi\!-\!\eta$ mixing to the singlet component of the
NN potential. We have found that the changes generated from off-shell mixing
were comparable to those observed when the point coupling was modified
by vertex corrections. Although off-shell mixing lead to a suppression of
the NN potential relative to the on-shell value, the changes were nowhere
as dramatic as those observed for $\rho-\omega$
mixing~\cite{goldman91,piewil93}.

These off-shell corrections to the NN potential should, however, be
taken with caution until additional momentum-dependent corrections are
investigated. For example, we have assumed that $N\bar{N}$ loops generate
a nonzero meson-mixing amplitude. These $N\bar{N}$ loops, however, are also
responsible for the vacuum dressing of the (unmixed) $\pi$ and $\eta$
propagators. These additional dressing should be studied for consistency
but might not be of practical importance.

In order to justify this last point we must go back the recurring, yet
little understood, topic of vertex corrections in hadronic field theories.
In principle, the calculation of (diagonal) self-energy corrections
to the $\pi$ and $\eta$ propagators could parallel the calculation
of off-diagonal ({\it e.g.,} $\pi\!-\!\eta$ or $\rho-\omega$) mixing.
In contrast to off-diagonal mixing, however, diagonal self-energy
corrections involve the sum, as opposed to the difference, of
proton and neutron loops. Thus, no small parameter (like
$\Delta M / M$ in the case of off-diagonal mixing) emerges in
the calculation of the unmixed propagators.
Consequently, one must use Dyson's equation to sum the lowest order
self-energy correction to infinite order. However, this procedure modifies
the analytic structure of the propagator and, because the structure of the
$NN-$meson vertex is typically neglected, generates
an unphysical ghost pole at spacelike momenta~\cite{chin77,furhor88}.

Recently, however, Allendes and Serot have computed the lowest order
self-energy correction to the $\omega$-meson propagator by including a
$NN\omega$ vertex approximated by its on-shell form~\cite{allser92}.
Solving Dyson's equation with this vertex-corrected self-energy resulted
in an $\omega$-meson propagator with no ghost poles and finite at large
spacelike momenta. Moreover, no significant changes in the low momentum
region (\hbox{$-q^{2}/M^{2}<1$}) were seen between the free and dressed
propagators. This last result is the only justification we have, so far,
in presenting the NN potential of Fig.~\ref{figthree}. How will
vertex corrections modify the pseudoscalar propagators and how sensitive
are these corrections to the off-shell behavior of the vertex are
important open questions that must be address before a clear picture
of the role of meson mixing on CSV observables will emerge.

\acknowledgments
This research was supported by the Florida State University
Supercomputer Computations Research Institute and U.S. Department
of Energy contracts DE-FC05-85ER250000, DE-FG05-92ER40750.

\figure{The $\pi\!-\!\eta$ mixing amplitude as a function of $q^{2}$
        with (dashed and dotted lines) and without (solid and dot-dashed
        lines) the inclusion of form factors in the spacelike region.
        The experimental-like symbols show results from a calculation
        by Maltman using chiral perturbation theory~\cite{maltman92}.
        \label{figone}}
\figure{The ratio of pseudoscalar ($\pi\!-\!\eta$) to vector
        ($\rho-\omega$) mixing amplitudes as a function of
         $q^{2}$ with (dashed line) and without (solid line)
         the inclusion of form factors in the spacelike region.
         Calculations were done with the parameters of
         Table~\ref{tableone}.
        \label{figtwo}}
\figure{The contribution from $\pi\!-\!\eta$ mixing to the NN potential
        as a function of $q^{2}$ using the off-shell value for the
        mixing amplitude (dot-dashed line), and the on-shell
        value with (dashed line) and without (solid line)
        the inclusion of on-shell form factors at the external
        nucleon legs. Calculations were done with the parameters
        of Table~\ref{tableone}.
        \label{figthree}}

 \mediumtext
 \begin{table}
  \caption{Meson masses, coupling constants, tensor-to-vector ratio
           and cutoff parameters. (see Table~4 of Ref.~\cite{machl87}
           and Eq.~(31) of Ref~\cite{coon82}).}
   \begin{tabular}{ccccc}
    Meson & Mass(MeV) & ${g^2/4\pi}$
          & $C=f/g$ & $\Lambda({\rm MeV})$ \\
        \tableline
    $\pi$     &  138  &  14.1 &  ---  &  1300  \\
    $\eta$    &  549  &  0.90 &  ---  &  1500  \\
    $\rho$    &  770  &  0.41 &  6.1  &  1400  \\
    $\omega$  &  783  &  10.6 &  0.0  &  1500  \\
   \end{tabular}
  \label{tableone}
 \end{table}

\end{document}